\documentstyle[aps,prl,epsf]{revtex}
\tightenlines

\newcommand{\be}{\begin{equation}}
\newcommand{\ee}{\end{equation}}
\newcommand{\bea}{\begin{eqnarray}}
\newcommand{\eea}{\end{eqnarray}}

\begin{document}

\draft


\title{Hybrid Quarkonia on Asymmetric Lattices}

\author{CP-PACS Collaboration\\
$^1$T.~Manke, $^1$H.~P.~Shanahan, $^1$A.~Ali Khan, $^2$S.~Aoki, 
$^{1,2}$R.~Burkhalter, $^1$S.~Ejiri,
$^3$M.~Fukugita, $^4$S.~Hashimoto, $^{1,2}$N.~Ishizuka, $^{1,2}$Y.~Iwasaki, 
$^{1,2}$K.~Kanaya, $^1$T.~Kaneko, $^4$Y.~Kuramashi, $^1$K.~Nagai,
$^4$M.~Okawa, $^{1,2}$A.~Ukawa, $^{1,2}$T.~Yoshi\'e}

\address{
$^1$Center for Computational Physics,
University of Tsukuba, Tsukuba, Ibaraki 305-8577, Japan \\
$^2$Institute of Physics, University of
Tsukuba, Tsukuba, Ibaraki 305-8571, Japan \\
$^3$Institute for Cosmic Ray Research,
University of Tokyo, Tanashi, Tokyo 188-8502, Japan \\
$^4$High Energy Accelerator Research Organization
(KEK), Tsukuba, Ibaraki 305-0801, Japan}

\date{\today}

\maketitle
\vspace*{-5.5cm}
\noindent
\vspace*{5.5cm}

\begin{abstract}
We report on a study of heavy quark bound states containing 
an additional excitation of the gluonic degrees of freedom.
To this end we employ
the NRQCD approach on coarse and asymmetric lattices, where we discard vacuum
polarisation effects and neglect all spin-correction terms.
We find a clear hybrid signal on all our lattices ($a_s= 0.15 \ldots 0.47$ fm).
We have studied in detail the lattice spacing artefacts,
finite volume effects and mass dependence.
Within the above approximations we predict the lowest lying 
hybrid excitation in Charmonium to be 1.323(13) GeV above the ground state,
where we use the 1P-1S splitting to set the scale. 
The bottomonium hybrid was found to be 1.542(8) GeV above its ground state.
\end{abstract}


Gluonic excitations are ideal objects for investigating
the nonperturbative nature of the gluonic degrees of freedom in QCD. 
Hybrid mesons can be thought of as hadronic bound states with an additional
excitation of the gluon flux.
This can give rise to states with non-conventional
quantum numbers and has triggered an intense experimental and theoretical
search for such particles.
Previous predictions for the energies of hybrid states come from 
phenomenological models \cite{phenom_hybrid}, static potential models 
\cite{cm_hybrid,morning_hybrid} and lattice simulations with propagating quarks
\cite{milc_hybrid,cm_light_hybrid,hybrid_lat97}. 
So far, lattice QCD is the only approach in which hybrids can be treated
from first principles. However, the errors from such calculations 
on isotropic lattices are still much larger than for conventional states.
This is because the correlation functions decay too rapidly when the
excitation energy is very large with respect to the inverse lattice spacing.
To obtain a similar signal-to-noise ratio one needs a much finer resolution
in the temporal direction.

While in recent years there has been much progress in obtaining more reliable 
results from improved actions on spatially coarse lattices, it has also been demonstrated
that anisotropic lattices can be employed to accommodate different physical
scales on the same lattice. In particular the study of glueball states on coarse and 
anisotropic lattices \cite{peardon_and_morning} has prompted us to study
heavy hybrid states on such lattices in order to increase both 
the scope and the precision of a previous calculation \cite{iso_hybrid_98} significantly. 
There a non-relativistic approach (NRQCD) was used for the heavy $b$ quarks 
on an isotropic lattice with  $a \approx 0.08$ fm.
NRQCD has frequently been employed to allow high precision measurements for the
$b\bar b$-system \cite{nrqcd_prec,omv6,spitz,alpha_nrqcd}.
Also the combination of improved gluon actions and the NRQCD
approach for heavy quarks has already been used to determine the spectrum of 
conventional quarkonia \cite{trottier,charm_coarse,ron_lat98}.
Previous attempts to measure heavy hybrid states 
on the lattice were reviewed in \cite{kuti_plenary}.
An initial preliminary study using asymmetric lattices for Bottomonium hybrids was presented in 
\cite{morning_hybrid,juge_etal}.


In this paper we implemented such an efficient approach to study in detail 
the lattice artefacts and finite size effects for both the $b\bar bg$ and 
$c\bar cg$ hybrid. For the latter state we obtain excellent agreement with a 
relativistic simulation on isotropic lattices \cite{milc_hybrid}. 
As with this other study we have neglected dynamical sea quark effects, but
we have succeeded in lowering the statistical error to about $1\%$.

In our study we generated the gauge field configurations using a
tadpole-improved action which has been employed by different 
groups \cite{peardon_and_morning,ron_lat98}:

\bea
S =  - \beta \xi^{-1} \sum_{x, {\rm i > j}} \left\{\frac{5}{3}P_{\rm i j} -
\frac{1}{12}\left(R_{\rm i j} + R_{\rm i j}\right)\right\}  
- \beta \xi \sum_{x,{\rm i}} \left\{\frac{4}{3}P_{\rm i t} -
\frac{1}{12}R_{\rm i t}\right\} ~~,~~ U_i \to U_i/u_s~~.
\label{eq:action}
\eea

\noindent Here $P_{\rm ij}$ and $R_{\rm ij}$ denote the trace of the standard spatial plaquette
and rectangle, respectively. Where the index $t$ appears the
plaquette/rectangle extends only one link into the temporal direction.
This theory has two parameters, $\beta$ and $\xi$, the second of which 
determines the asymmetry of our lattices. At tree level the ``aspect ratio''
is $\xi=a_s/a_t$, where $a_s$ and $a_t$ are the spatial and temporal lattice
spacings, respectively.
From \cite{peardon_and_morning} we note that the radiative
corrections to this relation are small when tadpole improvement is implemented,
as described below.

The action in Equation \ref{eq:action} is designed to be accurate up to ${\cal
O}(a_s^4,a_t^2)$ classically. To account for radiative corrections all spatial gauge 
links, $U_i$, are self-consistently tadpole-improved with 
$u_s=\langle 0|(1/3)~P_{\rm i j}|0 \rangle^{1/4}$, as suggested in \cite{viability_pt}. 
Such a mean-field treatment was demonstrated to reduce significantly the leading corrections,
which are due to unphysical tadpoles in lattice perturbation theory. 
Our scaling analysis shows that errors ${\cal O}(\alpha a_s^2)$ are
indeed negligible if the lattice spacing is sufficiently small. 
Since we will use this action only for small temporal lattice spacings
we expect ${\cal O}(\alpha a_t^2)$ errors to be very small. 
Therefore we have not employed the tadpole improvement description 
for the temporal gauge links.
Our results for two different aspect ratios $\xi$ justify such
an assumption.

To propagate the heavy quarks through the lattice we expand the NRQCD
Hamiltonian correct to ${\cal O}(mv^2)$, where all spin-dependent terms are absent.
This accuracy was already employed in \cite{iso_hybrid_98}. 
The only additional improvement is to correct for temporal and spatial lattice 
spacing errors by adding two extra terms ($c_7,c_8$) to the evolution
equation in \cite{omv6}:
\bea
H_0 = -  \frac{\Delta^2}{2m_b} ~~,~~ 
\delta H = - c_7 \frac{a_t\Delta^4}{16n m_b^2} + c_8 \frac{a_s^2 \Delta^{(4)}}{24m_b} ~~.
\label{eq:dh}
\eea


\noindent In those operators all spatial links are also tadpole-improved using
the same $u_s$ as for generating the configurations. After this modification
we take the tree-level values for all the coefficients in the Hamiltonian.
In this case, $c_7=c_8=1$.

For the non-relativistic meson operators we only used the simplest possible choices 
and have not tuned the overlap to optimise the signal. 
The gauge-invariant construction of S-state and P-state operators is described in \cite{omv6}. 
For the {\it magnetic} hybrid signal studied here, we have inserted the lattice version
of the magnetic field into the $Q\bar Q$-state ($B_i = \epsilon_{ijk}\Delta_j\Delta_k$):
\be
^1H_i(x) = \psi^{\dag}(x)B_i \chi(x) ~~,~~^3H_{jk}(x) = \psi^{\dag}(x)\sigma_jB_k\chi(x) ~~.
\label{eq:hybrid}
\ee
For the leading order in the NRQCD Hamiltonian, the operators in Equation
\ref{eq:hybrid} create a whole set of degenerate states: $1^{--},0^{-+},1^{-+},2^{-+}$.
These are the spin-singlet and spin-triplet states with zero orbital angular 
momentum, including the exotic combination $1^{-+}$. This degeneracy will be
lifted when higher order relativistic corrections are re-introduced into the
NRQCD Hamiltonian. We expect this to be a small effect, in as much as the
heavy quarks are very slow in the shallow hybrid potential
\cite{morning_hybrid}. By the same argument, we expect hybrid states with
additional orbital angular momentum to be almost degenerate as it was observed in \cite{iso_hybrid_98}.
The definition of the operators in Equation \ref{eq:hybrid} has been augmented 
by a combination of fuzzing \cite{fuzz} for the
links and Jacobi-smearing for the quark fields \cite{jacobi}. No effort has been made
to optimise the signal further, but one could do so if
even higher precision is needed, or if higher excited states are to be determined.
To extract the hadron masses we simply fit the meson correlators
to a single exponential;

\be
C_\alpha(t) =   \langle H_\alpha^{\dag}(t)H_\alpha (0) \rangle = A_{\alpha}e^{-m_{\alpha}t}~~.
\label{eq:fit}
\ee

\begin{figure}
\begin{center}
\leavevmode
\hbox{\epsfxsize = 7 cm \epsffile{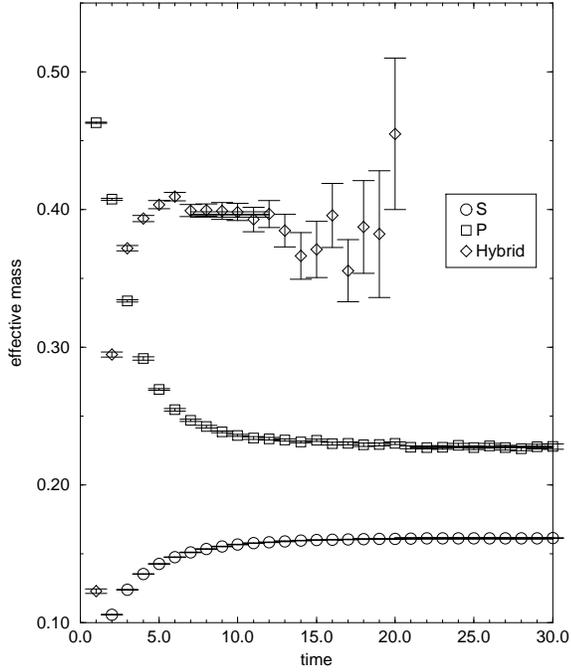}}
\end{center}
\caption{A representative effective massplot for the S, P and hybrid
state in Bottomonium at $(\beta,\xi) = (2.7,5), a_sm_b=3.15$ on a $9^3\times 40$
lattice.}
\label{fig:eff}
\end{figure}

We measured correlators every 10 trajectories in the Monte Carlo update, 
and for the error analysis we binned 50 such measurements into one. After this binning we still have an 
ensemble of 100--1000 configurations depending on the lattice and the state of interest.
As it can be seen from the representative example of an effective
massplots in Figure \ref{fig:eff}, the data is
very good and the goodness of the single exponential fits is
always bigger than $Q=0.1$, which we called acceptable.
In Tables \ref{tab:results_charm} and \ref{tab:results_ups} we present our results
and the simulation parameters.

To determine the lattice spacing $a_t^{-1}$, we used
the 1P-1S splitting in Charmonium and Bottomonium. As expected, the values from
Charmonium are smaller than those from Bottomonium, because in the quenched
approximation the coupling does not run as in full QCD.
At $(\beta,\xi)=(2.4,5)$ we observe a $16\%$ effect.
In order to give another estimate of quenching errors we also
determined the radial excitations, $nS$, and calculated the ratio 
$R_{\rm SP}$= 2S-1S/1P-1S, which can be compared with the experimental
value of 1.28 for Bottomonium. At $(\beta,\xi)=(2.7,5)$ we find $R_{\rm
SP}=1.424(89)$. From these findings we quote quenching errors of $10-20\%$.
Several suggestions have been made how to measure the spatial lattice
spacing \cite{ron_lat98,klassen}, but to convert our results into dimensionful 
numbers we can use $a_t^{-1}$ throughout. 

We have also tested the velocity expansion and included relativistic
corrections up to ${\cal O}(mv^6)$ for some of our lattices.
At this level of accuracy we could not resolve any significant change
in our results. A more detailed analysis of the spin structure is subject 
of a future project.

In this study we were mainly interested in the gluonic excitations
of heavy quarkonia. For this purpose it was irrelevant to adjust the 
quark masses to their exact values. 
For some lattices we have changed the quark masses by 25\% and did not
find any noticeable change in the ratio $R_{\rm H}$ = (1H-1S)/(1P-1S).

\begin{figure}
\begin{center}
\leavevmode
\hbox{\epsfxsize = 7 cm \epsffile{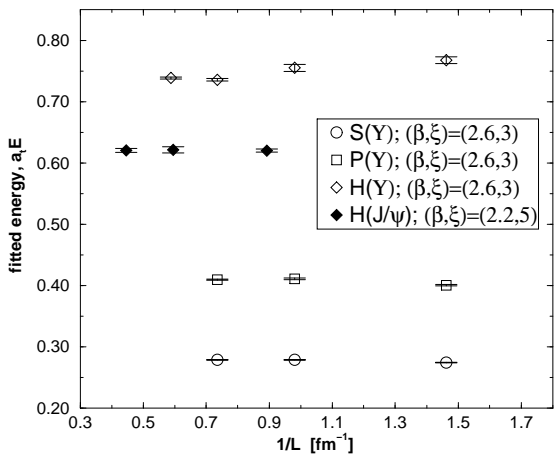}}
\end{center}
\caption{Finite volume analysis for Charmonium and Bottomonium. 
We plot the dimensionless energies for different states against
the inverse spatial extent, 1/L, of the lattice.}
\label{fig:finite}
\end{figure}

Finite volume effects were a source of immediate concern for us.
This is because hybrid states are expected to reside in a very flat potential
\cite{morning_hybrid}. The bag model also suggests a very large bound state as the
result of the gluonic excitation \cite{bag_hybrid}. As shown
in Figure \ref{fig:finite}, we have found that for spatial extents of 1.2 fm or larger
the masses of all Bottomonium and Charmonium states remain constant within small statistical
errors. For even smaller volumes we can resolve a slight increase in the mass of
the $b\bar bg$ hybrid. In other words, the spatial extent of the hybrid
excitation seems to be almost independent of the heavy quark mass.
In this sense the $b\bar bg$ hybrid is more difficult to calculate, as we need similar 
volumes but finer lattices than for Charmonium. 

\begin{figure}
\begin{center}
\leavevmode
\hbox{\epsfxsize = 7 cm \epsffile{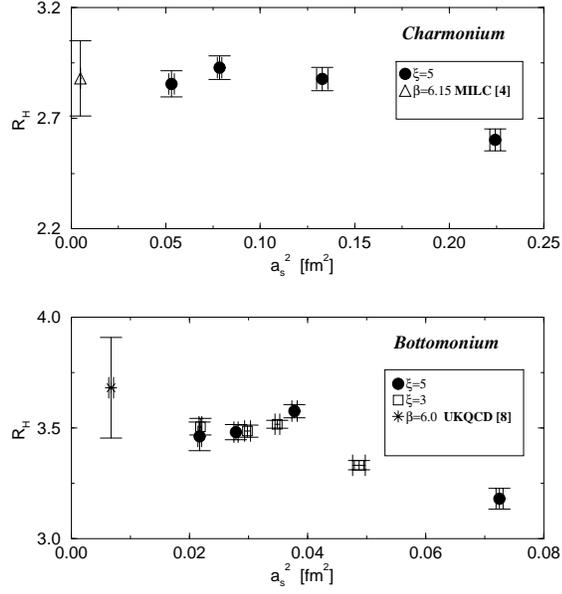}}
\end{center}
\caption{Scaling of the hybrid excitation, $H-S$. We plot the ratio $R_{\rm H} =
(H-S)/(P-S)$ against the squared spatial lattice spacing; $a_s = \xi a_t$ at
tree level. We also show the results from a previous calculation [8] on a
symmetric lattice at $\beta=6.0$ (burst). To display the result for the
lowest $c\bar cg$ hybrid at $\beta=6.15$ [4], we use their value 1.32(8) GeV and 
the experimental P-S splitting in Charmonium (triangle).}
\label{fig:scaling}
\end{figure}

Finally, we carried out a scaling analysis to demonstrate that discretisation errors
are under control. This is of utmost importance for the NRQCD
approach since one cannot extrapolate to zero lattice spacing in this effective
field theory and one has to model continuum behaviour already for finite lattice spacings.
In Figure \ref{fig:scaling} we show the scaling of the hybrid excitation above 
the ground state. From this, one can see that we have found convincing scaling windows for both
Bottomonium and Charmonium. Scaling violations can only be seen on the coarsest lattices
($a_s > 0.36$ fm for $c\bar cg$ and $a_s > 0.19$ fm for $b\bar bg$), 
but this is not totally unexpected - it is questionable how well our simple
minded implementation of the tadpole prescription works to remove the 
${\cal O}(\alpha p^2a^2)$ errors for heavy quark systems on such coarse lattices. 
In the case of $b\bar bg$ we also plot the results for two different aspect ratios.
Both results are consistent and confirm the initial assumption of 
small temporal lattice spacing errors. 
To quote our final result for the lowest lying hybrid excitations
we take the averaged value of all the results in the scaling region
and employ the experimental values for the 1P-1S splitting to set the scale.
We find 1.323(13) GeV for the case of Charmonium and 1.542(8) GeV for the 
first gluonic excitation in Bottomonium, in good agreement with a previous estimate 
of 1.68(10) GeV \cite{iso_hybrid_98}.

In conclusion, we have demonstrated the usefulness of coarse and anisotropic
lattices for the nonperturbative study of gluonic excitations in heavy quark systems. 
Furthermore, this should also be
considered a success of the NRQCD approach, which allowed
us to predict the lowest lying Charmonium hybrid state at the same mass as from
a relativistic calculation \cite{milc_hybrid}.
Apart from the very accurate predictions for hybrid quarkonia,
it is also interesting to notice that all of the above results could be obtained
in a comparatively short period of time. 
Whereas our present calculation confirms that the $b \bar bg$ hybrid will lie
above the S+S threshold for decay into B-mesons, the issue whether it may be found below
the S+P threshold has to be decided in a simulation where dynamical sea
quarks are included in order to control this last remaining systematic error.


TM would like to thank R.R. Horgan and I.T. Drummond
for many useful and constructive discussions.
The calculations were done using workstations and the CP-PACS facilities at
at the Center for Computational Physics at the University of Tsukuba.
This work is supported in part by the Grants-in-Aid of Ministry of
Education (No. 09304029). TM, HPS and AAK are supported by the 
JSPS Research for Future Program, and SE and KN are JSPS Research Fellows.


\widetext
\begin{table}
\begin{center}
\begin{tabular}{|l|l|l|l|l|l|}
\hline
($\beta,\xi$)  &  (1.7,5)         & (1.9,5)         & (2.2,5)          & (2.4,5)         & (2.4,5)       \\
Volume         &  $4^3 \times 40$ & $4^3 \times 40$ & $8^3 \times 40$  & $8^3 \times 40$ & $8^3 \times 40$ \\
$u_s$          &  0.7370          & 0.7568          & 0.7841           & 0.7997          & 0.7997       \\
$a_sm_c$       &  3.0             & 2.66            & 2.0              & 1.62            & 1.62, ${\cal O}(mv^6)$        \\
\hline                                        
\hline                                        
P-S            &   0.2196(18)     & 0.1689(26)      & 0.1299(13)       & 0.1068(21)      & 0.1047(42)      \\
$a_t^{-1}$/GeV &   2.084(18)      & 2.709(42)       & 3.522(37)        & 4.286(86)       & 4.37(18)        \\
\hline                                        
H-S            &  0.5713(98)      & 0.4860(48)      & 0.3821(33)       & 0.3048(17)      & 0.3011(24)       \\
H-S/P-S        &  2.602(49)       & 2.877(53)       & 2.928(53)        & 2.855(59)       & 2.87(12)              \\
H-S/GeV        &  1.191(23)       & {\bf 1.317(24)} & {\bf 1.346(18)}  & {\bf 1.306(27)} & 1.316(54)  \\
\hline
\end{tabular}
\caption{Results for Charmonium. The dimensionful numbers in the scaling region are given 
in boldfaced characters. From their average we obtain {\bf 1.323(13)} GeV
for the lowest lying hybrid excitation from our simulation with accuracy
${\cal O}(mv^2,a_s^4,a_t^2)$. 
In the last column we give the spin averaged results from a
higher order accuracy ${\cal O}(mv^6,a_s^4,a_t^2)$.}
\label{tab:results_charm}
\end{center}
\end{table}

\begin{table}
\begin{center}
\begin{tabular}{|l|l|l|l|l|l|l|l|}
\hline
($\beta,\xi$)  & (2.4,5)        & (2.6,5)        & (2.7,5)        & (2.7,5)        & (2.5,3)        & (2.6,3)        & (2.8,3) \\       
Volume         & $8^3\times 40$ & $6^3\times 40$ & $9^3\times 40$ & $9^3\times 40$ & $6^3\times 40$ & $8^3\times 40$ & $10^3\times 40$ \\
$u_s$          &  0.7997        & 0.8139         & 0.8193         & 0.8193         & 0.8100         &  0.8193        &  0.8314        \\
$a_sm_b$       &  4.73          & 3.50           & 3.15           & 3.15, ${\cal O}(mv^6)$            & 4.10           &  4.00          &  3.33          \\
\hline                                                                                     
\hline                                  
 P- S          & 0.08665(68)    & 0.07444(65)    & 0.06573(62)    & 0.0636(23)     & 0.1388(10)     & 0.13068(82)    & 0.10988(77)    \\
$a_t^{-1}$/GeV & 5.075(43)      & 5.907(54)      & 6.690(67)      & 6.91(25)       & 3.169(25)      & 3.365(23)      & 4.002(30)  \\
\hline                                                                                                               
H-S            & 0.30988(84)    & 0.2591(11)     & 0.2275(36)     & 0.2305(46)     & 0.4882(41)     & 0.4568(21)     & 0.3852(32)    \\
H-S/P-S        & 3.576(30)      & 3.481(34)      & 3.462(65)      & 3.62(15)       & 3.517(39)      & 3.496(27)      & 3.506(38)     \\
H-S/GeV        & {\bf 1.573(14)}& {\bf 1.531(15)}& {\bf 1.522(29)}& 1.594(67)      & {\bf 1.547(18)}& {\bf 1.537(13)}& {\bf 1.542(17)}    \\
\hline
\end{tabular}
\caption{Results for Bottomonium. From the average over the scaling region 
we obtain {\bf 1.542(8)} GeV for the lowest lying $b\bar bg$ hybrid, when the
P-S splitting is used to set the scale. In column 5 we show results with
accuracy ${\cal O}(mv^6,a_s^4,a_t^2)$.}
\label{tab:results_ups}
\end{center}
\end{table}

\narrowtext

\end{document}